\documentclass[letterpaper, 10 pt, conference]{ieeeconf}  

\IEEEoverridecommandlockouts                              
\usepackage[utf8]{inputenc}
\usepackage[T1]{fontenc}
\usepackage{graphicx}
\usepackage{xcolor}
\usepackage{acronym}
\usepackage{graphicx} 
\usepackage{booktabs} 
\usepackage{xparse}   
\usepackage{amsmath}
\makeatletter
\let\NAT@parse\undefined
\makeatother
\usepackage[bookmarks,colorlinks,citecolor=green]{hyperref}
\usepackage{bbold}

\acrodef{eeg}[EEG]{electroencephalogram}
\acrodef{hmm}[HMM]{hidden Markov model}
\acrodef{dl}[DL]{deep learning}
\acrodef{dnn}[DNN]{deep neural network}
\acrodef{cnn}[CNN]{convolutional neural network}
\acrodef{dwt}[DWT]{discrete wavelet transform}
\acrodef{rnn}[RNN]{recurrent neural network}
\acrodef{flops}[FLOPs]{floating point operations}
\acrodef{lstm}[LSTM]{Long-Short Term Memory}
\acrodef{mi}[MI]{Mutual Information}

\title{\LARGE \bf
 Efficient  Epileptic Seizure Detection Using CNN-Aided Factor Graphs
}

\author{Bahareh Salafian$^{1}$, Eyal Fishel Ben$^{2}$, Nir Shlezinger$^{2}$, Sandrine de Ribaupierre$^{1,3}$, and Nariman Farsad$^{4}$ 
\thanks{$^{1}$B. Salafian is with the School of Biomedical Engineering, University of Western Ontario, London, ON N6A 5B9
        {\tt\small bsalafia@uwo.ca}}%
\thanks{$^{2}$E. Fishel and N. Shlezinger are with the School of Electrical and Computer Engineering, Ben-Gurion University of the Negev, Be'er Sheva, Israel, 84105
        {\tt\small \{eyalfish@post.bgu.ac.il; nirshl@bgu.ac.il\}}}%
\thanks{$^{3}$S. de Ribaupierre is with Department of Clinical Neurological Sciences and the School of Biomedical Engineering, University of Western Ontario, ON N6A 5B9
        {\tt\small sderibau@uwo.ca}}%
\thanks{$^{4}$N. Farsad is with the  Department of Computer Science at Ryerson University,Toronto, ON M5B 2K3
        {\tt\small nfarsad@ryerson.ca}}%

}

\begin{document}

\maketitle
\thispagestyle{empty}
\pagestyle{empty}

\begin{abstract}
We propose a computationally efficient algorithm for seizure detection. Instead of using a purely data-driven approach, we develop a hybrid model-based/data-driven method, combining convolutional neural networks with factor graph inference. On the CHB-MIT dataset, we demonstrate that the proposed method can generalize well in a 6 fold leave-4-patient-out evaluation. Moreover, it is shown that our algorithm can achieve as much as 5\% absolute improvement in performance compared to previous data-driven methods. This is achieved while the computational complexity of the proposed technique is a fraction of the complexity of prior work, making it suitable for real-time seizure detection.
\end{abstract}

\section{Introduction}
Epilepsy is one of the most common neurological disorders affecting about 50 million people worldwide. This disorder is associated with recurrent episodes of abnormal neural activity in the central nervous system known as epileptic seizures~\cite{noauthor_epilepsy_nodate}. Based on the area where seizure starts and the intensity of brain's abnormal signals, patients with epilepsy may suffer from different symptoms including auras, muscle contraction, and loss of consciousness~\cite{park_epileptic_2018}. Epilepsy affects the patients' private and professional life; for instance, some activities such as swimming, bathing, and climbing a ladder, become dangerous as a seizure during that activity might result in unpredictable injuries, and even death. Therefore, early detection of epilepsy can notably improve the patient's quality of life. 

The most common tool used to diagnose seizures is  \ac{eeg}~\cite{zazzaro_eeg_2019}. While other various techniques, such as magnetic resonance imaging (MRI)~\cite{kulaseharan_identifying_2019}, magnetoencephalography (MEG)~\cite{van_klink_simultaneous_2019}, and positron emission tomography (PET)~\cite{pianou_imaging_2019} are sometimes used in conjunction to \ac{eeg}, \ac{eeg} is widely preferred as it is economical, portable, non-invasive and shows clear rhythms in the frequency domain~\cite{subasi_epileptic_2019}. However, the review of \ac{eeg} signals is a time-consuming process, as a neurologist needs to monitor the recording. Expertise is needed to diagnose epilepsy as each case is quite variable, with different channels involved, while the spectral content of the rhythmic activity varies across individuals and signals, and is mostly contaminated by physiologic and non-physiologic interference~\cite{golmohammadi_deep_2017}. Hence, automatic seizure detection is a valuable clinical tool to address this issue and reduce the dependency on human~experts. 

Many machine learning studies have been developed for automatic seizure detection problems. One of the most common approaches  applies a support vector machine (SVM), which is mainly followed by additional preprocessing steps such as \ac{dwt} and fast Fourier transform (FFT) to extract more features of \ac{eeg} signals~\cite{slimen_eeg_2020,ahmad_prediction_2020,raghu_automated_2020,li_seizure_2021}. In the past decade, \ac{dl} techniques have become very popular in various applications, including the analysis of time series \ac{eeg} signals. Therefore, different \ac{dl} models have been investigated and tested in the area of seizure detection. For instance, Khalilpour et al.~\cite{khalilpour_application_2020} applied a 1D \ac{cnn} to \ac{eeg} signals of five patients to predict preictal and interictal states of the brain. In~\cite{jana_1d-cnn-spectrogram_2020}, the spectrogram of \ac{eeg} measurements was used as input to a 1D \ac{cnn}. The authors in \cite{sharan_epileptic_2020} utilized \ac{dwt} to represent the \ac{eeg} segments, which is used as input to a 1D \ac{cnn}. Moreover, they combine the \ac{dwt} of the current, previous, and next block for predicting the label of the current block, which helps exploit the temporal correlations. Another common \ac{dl} architecture is 2D \ac{cnn}. Boonyakitanont et al.~\cite{boonyakitanont_comparison_2019} applied 2D \ac{cnn} to 24 epileptic recordings from 23 patients, where the signals are segmented into 4-second blocks. They showed state-of-the-art performance in terms of detection accuracy. 

Most of these prior techniques divide the \ac{eeg} recording into blocks and treat these blocks independently. This does not take advantage of the temporal correlations that exists between consecutive blocks. While there are methods based on \ac{cnn}-\ac{rnn} architectures~\cite{roy18GRU, liang_scalp_2020} that can mitigate this issue, it is well known that \ac{rnn}s have high computational complexity for training. Another method used for exploiting temporal correlations is \ac{hmm} \cite{lee_classification_2018}. \acp{hmm} belong to family of factorizable joint distributions which admit low-complexity inference via factor graph methods \cite{loeliger2004introduction}. Despite proliferation of seizure detection algorithms, having a computationally efficient algorithm that can generalize to different patients and perform seizure detection reliably in a real-time manner is lacking.

In this work, we propose a computationally efficient epileptic seizure detection algorithm based on a hybrid model-based/data-driven approach using \ac{cnn}-aided factor graphs. First, we carefully design a 1D CNN for estimating the probability that a 4-second block of \ac{eeg} is a seizure block. Our goal is to design a network that is applied to the \ac{eeg} signals directly, without feature engineering using transforms such as \ac{dwt} or FFT. When using such feature engineering, one must carefully select the parameters of the transforms, and the performance of the trained model can vary considerably based on these parameters. Moreover, while these transforms typically provide information about the frequency component of the \ac{eeg} signals, they are not necessarily impactful for capturing the dependence of the \ac{eeg} signals across different channels. Such a dependence can be indicative of epileptic seizure. Our proposed 1D CNN is designed to be able to capture the long term dependence between \ac{eeg} channels and operates on the signals directly with minimal processing. To exploit the temporal correlation between consecutive blocks for further improvement, we use factor graph inference, \textcolor{black}{specifically using \ac{hmm} models to capture temporal correlations among the signals.}  
Our proposed hybrid method is highly efficient compared to previous works, where we reduce the inference complexity by a factor of 2. Despite this decrease in computation, our method achieves up to 5\% absolute improvement in performance measures such as precision, recall, and F1-score in a 6-fold leave-4-patients-out evaluation.

The rest of this paper is organized as follows. In Section~\ref{sec:background} we describe the problem statement, the dataset that is used for model development and evaluation, and the baseline models. Then, in Section~\ref{sec:Method} we describe our proposed method, which is evaluated and compared to prior methods in Section~\ref{sec:results}. Finally,  Section~\ref{sec:conclusion} provides concluding remarks.

\section{Background, Dataset, and Baseline Models}
\label{sec:background}
In this section, we first discuss the seizure detection problem. We then describe the dataset that is used for our hybrid model-based/data-driven algorithm development and evaluation. We conclude this section by presenting the baseline methods that are used for comparison in this work.

\subsection{Seizure Detection Using EEG Signals}
\ac{eeg} is the electrical recording of the brain activities which is the most popular diagnostic and analytical tool for epileptic seizures. In seizure detection problem, there are some basic terms as follows: 
\begin{itemize}

\item Ictal state which is the time when the seizure occurs (from start to end). 

\item Preictal state is a period of time just before a seizure occurs. 

\item Interictal state refers to the period between seizures.

\item Postictal state is a period of time just after the seizure ends.

\end{itemize}
During reading an EEG, experts don't only look at seizures which are not always occurring when the EEG is done, specifically with short EEG recordings, but they also have to interpret interictal EEG signal and therefore, they are trying to detect interictal epileptiform discharges (IED)\cite{emmady_eeg_2020}. IEDs are generated by the synchronous discharges of a group of neurons in a region referred to as the epileptic focus; however, the detection of these spikes is difficult to accomplish due to their similarity to waves that are part of normal EEG or artifacts and the wide variability in spike morphology and background between patients. For instance, abnormalities like breach rhythm (normal rhythm seen with skull defects) can have focal, sharply contoured morphology and they might be inferred as epileptic seizures. As such, having an automatic system to detect and predict seizures will resolve these issues.

\subsection{Data Description}
\label{sec:Data}
In this section we detail the data used in our study of hybrid model-based/data-driven epileptic seizure detection. We first describe the raw \ac{eeg} data, 
after which we describe the pre-processing carried out prior to its usage for training and inference. 

\subsubsection{\ac{eeg} Data}
\label{ssec:DataEEG}
The dataset used in this study is the public CHB-MIT Scalp \ac{eeg}
Database collected at the Children’s Hospital Boston and consists of \ac{eeg} recordings from pediatric subjects with intractable seizures \cite{goldberger_physiobank_2000}\footnote{This database is available online at PhysioNet (\url{https://physionet.org/physiobank/database/chbmit/})}. Recordings were collected from 23 subjects: 5 males aged 3-22 years, 17 females aged 1.5-
19 years, and one anonymous subject. Each case contains  9 to 42 continuous EDF files from a single subject. The duration of the recordings in each file varies between one to four hours. All signals were sampled at frequency of  256 Hz with 16-bit resolution. 
Since case 21 was obtained 1.5 years after case 1 from the same female subject, we consider case 21 as a separate patient; therefore, our experiment includes 24 subjects.  Please note that since we are evaluating the algorithms using a 6-fold leave-4-patients-out method, this might negligibly effect only one of the folds.

\subsubsection{Data Pre-Processing}
\label{ssec:DataPre}
Since many of patient recordings do not contain any seizures, in order to have a more balanced samples from seizures, for each patient, we only selected EDF files that have at least one seizure. Moreover, since the length of the seizures are very short (from 7 seconds to 753 seconds) compared to the overall recording (from 959 seconds to 14427 seconds), we shorten the recording to 3 times the seizure duration before and 3 times the seizure duration after the seizure.  Therefore, for every second of seizure data, there are 6 seconds of non-seizure data. 

From the EEG channels, we use the 18 bipolar montage: FP1-F7, F7-T7, T7-P7, P7-O1, FP1-F3, F3-T3, T3-P3, P3-O1, FP2-F4, F4-C4, C4-P4, P4-O2, FP2-F8, F8-T8, T8-P8, P8-O2, FZ-CZ, CZ-PZ. A notch filter is used to remove 60 Hz line noise from each \ac{eeg} signal. Then 4-second blocks, with 1024 sample points per block, are used as moving window with step size of 1 second. The value of 4 seconds was chosen to provide a good trade off between the number of samples in a block and the stationarity of the observed signals over a block\cite{klonowski2009everything}. We have observed that when the width of window is increased, the seizure detection procedure is not accurate enough. We now describe some of the prior seizure detection algorithms developed using this dataset, which will be used as baselines in this paper. 


\subsection{Baseline Methods}
\label{sec:MethodBase}
We consider two recent works that presented state-of-the-art performance on the CHB-MIT dataset as baselines in these papers \cite{boonyakitanont_comparison_2019, gomez_automatic_2020}. The method in~\cite{boonyakitanont_comparison_2019} is designed to take the \ac{eeg} signals as input without applying any type of transforms. \textcolor{black}{Therefore, we employ the same structure used in~\cite{boonyakitanont_comparison_2019} where the input shape for the model is (18,1024,1), which implies considering each block as an image with the size of (18,1024) and channel dimension of 1.} The method in \cite{gomez_automatic_2020} takes a graphical image of the \ac{eeg} recording as input rather than the \ac{eeg} measurements. This type of image-based feature was shown to achieve better detection performance compared to other features-based methods in seizure detection such as spectrogram or periodogram in a recent study~\cite{cho_comparison_2020}.

\section{Methodology}
\label{sec:Method}

Our hybrid model-based/data-driven algorithm  combines a carefully designed \ac{cnn}, which estimates the presence of seizure in a 4-second block, with factor graph inference to exploit the temporal correlation between the blocks. We begin this section by describing the the \ac{cnn} architecture, followed by the factor graph based inference step.

\subsection{1D \ac{cnn} Architecture} 
2D \acp{cnn}, as utilized in the baseline methods detailed in Subsection~\ref{sec:MethodBase}, exploit the notion of {\em locality}, exhibited by natural images, which implies that the level of correlation between different elements typically grows the closer they are in the image. However, in \ac{eeg} segment matrices, each element represents a single \ac{eeg} measurement, which is likely to be correlated with all the remaining measurements taken at that time instance, regardless of their row index in the matrix representation. This structure makes 1D \acp{cnn}, which combine all measurements taken from different channels at a given time instance, more suitable compared to 2D \acp{cnn} used in~\cite{boonyakitanont_comparison_2019}. That allows the network to better learn the correlations that may exist between different channels since during seizures, the channel measurements can become highly correlated. 

\begin{figure}
    \centering
    \includegraphics[width=3.4in]{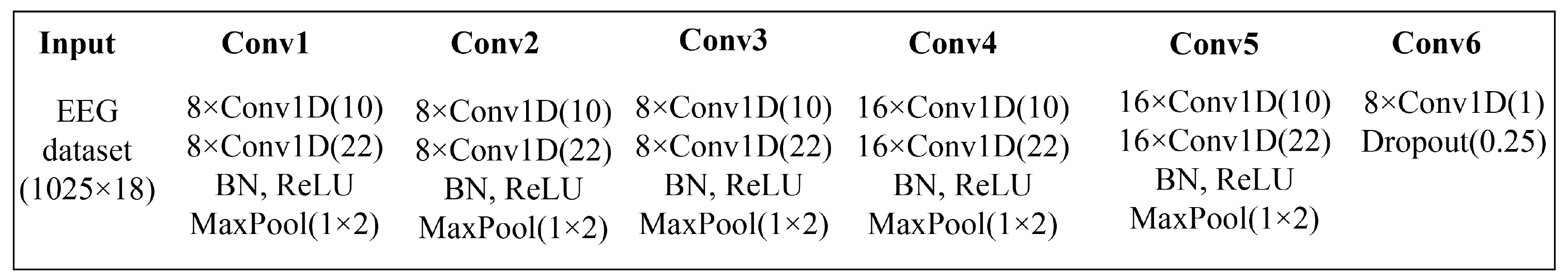}
    \caption{Proposed 1D \ac{cnn} architecture.}
    \label{fig:1DCNNV1}
\end{figure}

In order to have a comparable configuration with the baseline models, we use the same number of layers \textcolor{black}{where the input shape for this model is (1024,18)}. The inputs to our 1D \ac{cnn} are \ac{eeg} signals with minimal prepossessing, which only removes the 60 Hz component of the signals. The baseline models detailed in Subsection~\ref{sec:MethodBase} use a kernel size of 3 and 2. Given the number of layers, this results in a receptive field of approximately 30 milliseconds. Since this is not enough to capture low frequency components of the signal as well as the long term correlations between the \ac{eeg} channels, we design out kernel size to be much larger, which results in a receptive field that covers approximately 1 second of the data. Fig.~\ref{fig:1DCNNV1} shows the complete architecture of the proposed 1D \ac{cnn}.

As will be shown in Section \ref{sec:results}, these simple changes, namely choosing the kernel size carefully and using a 1D \ac{cnn}, improve the results significantly compared to the baseline \ac{cnn} model in \cite{boonyakitanont_comparison_2019}. This is because the network can capture a wider range of frequency components in the signal and also better capture the correlations between the \ac{eeg} channels. An additional advantage of using 1D \acp{cnn} stems from their reduced complexity during inference compared to their 2D counterparts. Although we have increased the receptive field of the network by increasing the kernel size for our proposed 1D \ac{cnn} architecture compared to \cite{boonyakitanont_comparison_2019}, the number of \ac{flops} during inference (i.e., the number of floating point multiplication and summation operations) for the proposed method is almost down by a factor of 2 compared to \cite{boonyakitanont_comparison_2019} and by a factor of 20 compared to \cite{gomez_automatic_2020} as summarized in Table~\ref{tab:numbers}. For training all networks, we use the ADAM optimizer with the learning rate of 0.001, batch size of 128 and 10 epochs of training. 

\begin{table}
\caption{Computational complexity in FLOPs for all models}
\label{tab:numbers}
\begin{center}
\begin{tabular}{|c||c|}
\hline & Mega FLOPs\\
\hline
2D \ac{cnn} \cite{boonyakitanont_comparison_2019} & 14.5  \\
\hline
2D \ac{cnn} \cite{gomez_automatic_2020} & 200  \\
\hline
1D \ac{cnn} & 9.81 \\
\hline
1D \ac{cnn}+FG & 9.81 \\
\hline
1D \ac{cnn}+GRU & 29.4 \\
\hline
\end{tabular}
\end{center}
\end{table}


\subsection{Factor Graph Based Inference} 
The 1D \ac{cnn} model outputs an estimate of the probability that a given 4-seconds block corresponds to a seizure. This probability is based solely on its input \ac{eeg} segment block, and does not account for the fact that the presence of a seizure in a given block is likely to also reflect on its preceding and subsequent blocks. To incorporate this temporal correlation, we combine the probability estimates over multiple blocks by assuming that the underlying temporal correlation can be represented using a factor graph, and utilize the sum-product method for inference \cite{kschischang2001factor}. In the following we first describe how the underlying dynamics of the seizure detection setup can be represented as a factor graph, after which we discuss the sum-product algorithm and elaborate on its combinations with the proposed 1-D \ac{cnn} model.

\subsubsection{Factor Graph Representation of Underlying Dynamics}
Factor graphs provide a visual representation of a multivariate function, typically a joint distribution measure, which can be factorized into a partition of local functions \cite{loeliger2004introduction}. These partitions capture the inherent statistical relationship among variables affecting each partition. 
To represent a multivariate function as a factor graph, every partition and every variable must be associated with a unique node. Edges connect function nodes to variables nodes if and only if the function is explicitly dependent on the corresponding variable. We adopt the Forney style representation of factor graphs, where variable nodes are replaced by edges \cite{forney}. This graphical representation  enables desired quantities to be computed at reduced complexity via message passing over the factor graph \cite{loeliger2004introduction}.

\begin{figure}
    \centering
    \includegraphics[width=3.5in]{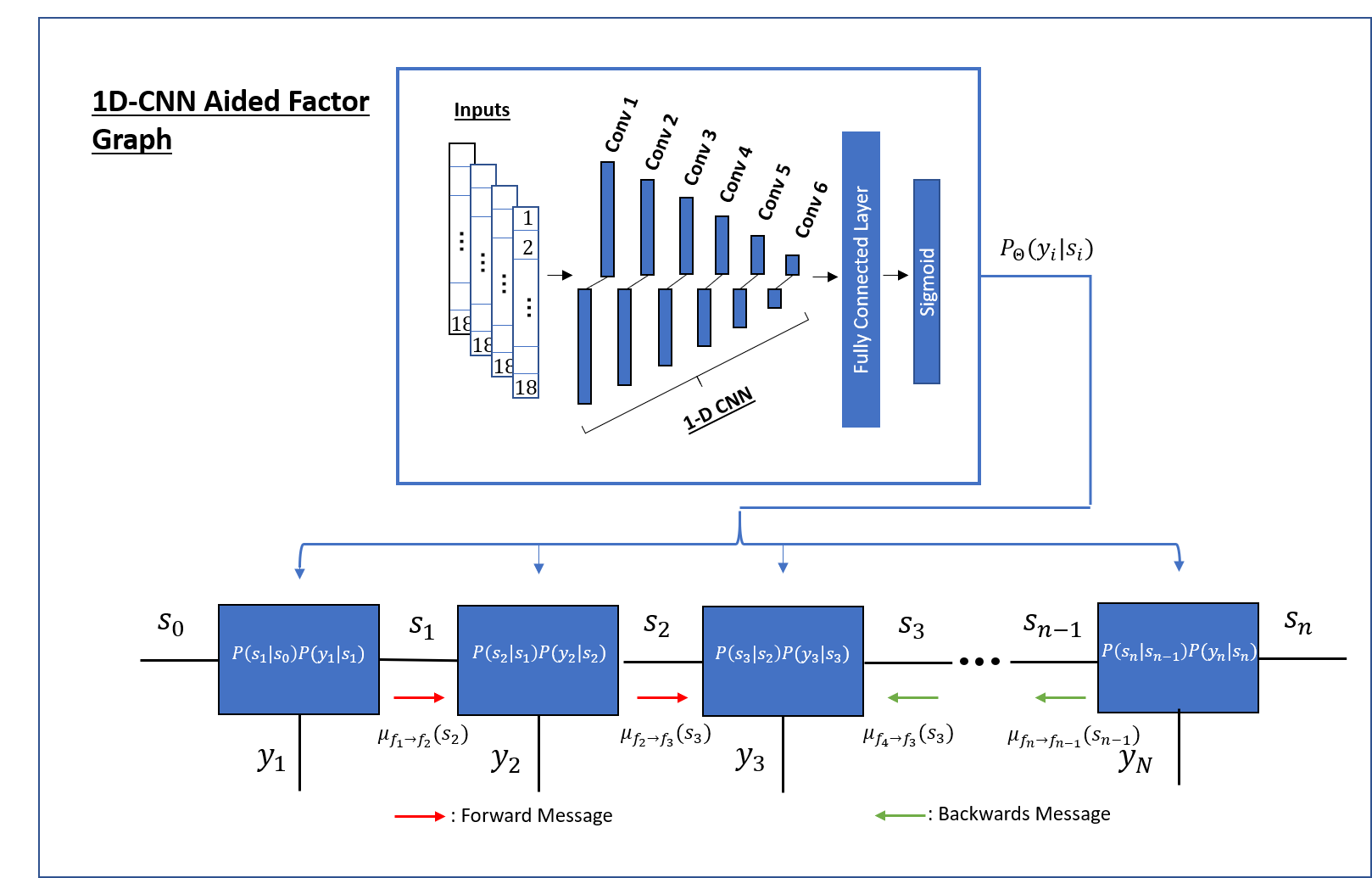}
    \caption{Factor Graph of a \ac{hmm} with function nodes computed using a 1-D \ac{cnn}. }
    \label{fig:FGarchitecture}
\end{figure}

To implement factor graphs inference, we first fix the structure of the graph, i.e., the interconnection between its nodes, which encapsulates our knowledge on the underlying statistical relationships for the seizure detection problem. A key feature preceding epileptic seizures is the de-synchronization of its rhythmic activity \cite{mormann_epileptic_2003}. To capture this behavior in our model study, we adopt a first-order \ac{hmm}. The Markovian architecture focuses on the temporal characteristics associated with the epileptic episodes, highlights the effects of temporal correlations in seizure detection, and facilitates efficient classification at reduced complexity. The resulting factor graph under this model is illustrated in Fig.~\ref{fig:FGarchitecture}.

To formulate this mathematically, let $\boldsymbol{y}=\{y_{1},y_{2},\cdots,y_{N} \}$, and $\boldsymbol{s}=\{s_{1},s_{2},\cdots,s_{N} \}$ describe the observed \ac{eeg} measurements and latent seizure states, respectively, over $N$ consecutive 4-seconds blocks. The latent states takes binary values, i.e., $s_{i} \in \{0,1\}$ corresponding to the presence or absence of a seizure. We assume that these states satisfy the Markovian property, and use  $P(s_{i} | s_{i-1})$ to denote the transition probability. The transition probability is taken as a control parameter, whose entries are fixed handcrafted values. In particular, we set $P(s_{i}=1 | s_{i-1}=0)$ to correspond to a $10.46\%$ of switching from non-seizure to seizure, and $17.9\%$ for transitioning in the opposite direction. \textcolor{black}{Ideally, one would like the transition probability matrix to reflect the true transitions probabilities between seizure and non-seizure state. However, the negligible seizure occurrences throughout the recording of EEG episodes, produces a highly unbalanced transition probabilities. Empirically, such transition probabilities did not facilitate accurate inference, hence hyper-parameter optimization of these probabilities was adopt, so that superior performance can be achieved.} To close, we assume each \ac{eeg} measurement  depends only on its corresponding seizure state. 
Under this postulated \ac{hmm}, the joint probability density function of the measurements and the states can be written as:
\begin{equation}
    \label{eqn:factor}
	P(\boldsymbol{s},\boldsymbol{y}) =  \prod_{i = 1}^{N} P(s_{i}|s_{i-1}) P(y_{i} | s_{i}).
\end{equation}

\subsubsection{The Sum-Product Algorithm} 
Having established the mathematical foundation and suitable factor graph representation, the objective is to distinguish between seizure and non-seizures states. Classification of these states is achieved through accurate inference of the marginal distribution $P(s_{i}, \boldsymbol{y})$, which is the metric used to compute the maximum a-posteriori probability detector. In principle, evaluating $P(s_{i}, \boldsymbol{y})$ from \eqref{eqn:factor} involves {\em marginalization}; a task whose computational burden scales exponentially with $N$. However, by employing factor graph inference via the sum-product algorithm, the same computation scales only linearly with $N$, making this operation computationally feasible for the problem at hand. 

The sum-product method relates the desired marginal probability to a product of "messages", where for each $k \in \{1,\ldots,N\}$ we write
\begin{equation}
\label{eqn:msgs}
    P(s_{k},\boldsymbol{y}) = \mu_{f_{j}\rightarrow s_{k}}(s_{k})\cdot \mu_{f_{j+1}\rightarrow s_{k}}(s_{k}).
\end{equation}
 The terms $\mu_{f_{j}\rightarrow s_{k}}(s_{k}), \mu_{f_{j+1}\rightarrow s_{k}}(s_{k})$ in \eqref{eqn:msgs} are interpreted as the "forward message" and "backward message" respectively, and are computed by \cite{loeliger2004introduction}
\begin{equation}
\label{eqn:ForwardPass}
\mu_{f_{j}\rightarrow s_{k}}(s_{k})=\sum_{\{s_{1},\cdots,s_{k-1} \}}\prod_{i = 1}^{n}f_{i}(y_{i},s_{i},s_{i-1}),
\end{equation}
and
\begin{equation}
\label{eqn:BackwardPass}
\mu_{f_{j+1}\rightarrow s_{k}}(s_{k})=\sum_{\{s_{k+1},\cdots,s_{N} \}}\prod_{i = n+1}^{N}f_{i}(y_{i},s_{i},s_{i-1}),
\end{equation}
where,
\begin{equation}
\label{eqn:functionNodes}
 f_{i}(y_{i},s_{i},s_{i-1})= P(s_{i}|s_{i-1}) P(y_{i} | s_{i}). \end{equation}


For a  Markov chain factor graph structure as in Fig.~\ref{fig:FGarchitecture}, the computational complexity is comprised of evaluating the forward \eqref{eqn:ForwardPass} and the backward \eqref{eqn:BackwardPass} messages. The recursive nature of the computations implies that the number of \ac{flops} grows linearly with the number of $4$-second blocks in a given \ac{eeg} episode, denoted by $N$. At the same time, the number of \ac{flops} also increases with the cardinality of the seizure state, denoted by $|\mathcal{S}|$, as well as the order of the Markov chain. Here, all seizure states are binary, hence $|S| = 2 $, and the factor graph is that of a first-order Markov chain. As a result, each message computation requires 4 multiplications and 2 addition operations. The result of these computations over a complete EEG episode requires merely $12N$ \ac{flops}, which is negligible compared to the complexity of applying the neural network models, as reported in Table~\ref{tab:numbers}. 
\subsubsection{\ac{cnn}-Aided Factor Graphs}
Implementing the sum-product algorithm requires knowledge of the underlying probability distribution $P(y_{i}|s_{i})$. In the case of seizure detection from \ac{eeg} measurements, acquiring such a distribution from first principles is unattainable. Accurately characterizing epileptic seizures to high fidelity requires a combination of dynamical features and spatial-temporal features, for example the entropy, amplitude, synchronization, spectral power amongst other features associated with epileptic episodes, making the underlying distribution highly complex and intractable. To contend with this difficultly, we follow the work \cite{shlezinger2020inference}, combining \ac{dl} models and inference algorithm.
In particular, we utilize the output of the 1D \ac{cnn} as an estimate of the conditional distribution $P(y_{i} | s_{i})$ required in order to compute the function nodes\footnote{In principle, a \ac{cnn} classifier is trained with the cross-entropy loss to detect $s_i$ from $y_i$, which is an estimate of the conditional distribution $P(s_{i}|y_{i})$. Here we use the \ac{cnn} output as an estimate of the conditional $P(y_{i} | s_{i})$, instead of converting it via Bayes rule as done in \cite{shlezinger2020viterbinet}, to avoid instabilities due to dividing by the estimated marginal of the seizure state~$s_i$.}. via \eqref{eqn:functionNodes}. 
Combining these approaches yields a hybrid model-based/data-driven detector, in which factor graph inference constitutes a robust final stage incorporating temporal correlation with the \ac{cnn} outputs as illustrated in Fig.~\ref{fig:FGarchitecture}.  

To complete the picture, the detection mechanism requires a comparison measure, i.e., a threshold, which represents a decision boundary for one to distinguish between different seizure states based on the estimated marginal distribution produced by the sum-product method.  Treating the threshold as fixed, if the resultant probability of a seizure state yielded by the \ac{cnn}-aided factor graph  exceeds that of the applied threshold, then the system is said to be in a seizure state. Otherwise the system is said to be in a non-seizure state. As shown in the sequel, this approach allows to achieve accurate detection with controllable tradeoffs between detection and false alarm. 



\section{Results and Discussion}
\label{sec:results}
We now discuss our evaluation results in this section. We begin by defining the performance metrics we employed for evaluation followed by the results and the discussions.

\subsection{Evaluation Method and Performance Metrics}
We apply 6-fold leave-4-patient-out cross validation to evaluate the performance and the generalizability of our hybrid model-based/data-driven approach. Specifically, in each fold, 4 patients are kept for testing and 20 patients for training. \textcolor{black}{The dataset used for evaluation, similar to train data, is segmented into 4-second blocks with moving window of 1 second where each block is labeled.} To evaluate the performance of the models trained in each fold, we use the following metrics.
\begin{itemize}
\item \textit{F1-score}: is a harmonic mean of recall and precision where former indicates the proportion of real positive cases that are correctly
predicted positive and latter denotes the proportion of predicted
positive cases that are correctly real positives \cite{powers_evaluation_2020}. 

\item \textit{AUC-PR}: is area under the precision-recall curve where high area under the curve means low false positive rate and low false negative rate.

\item \textit{AUC-ROC} is area under receiver operating characteristics (ROC) curve which is performance measurement for classification problems at various threshold settings. 
\end{itemize}
We choose these performance measures since the data is imbalanced (i.e., for every block with seizure we have 6 blocks with no seizures, and AUC measures are great indicator of performance over imbalanced data.

\begin{table}
\caption{Summary of Results}
\vspace{-0.8cm}
\label{tab:summary}
\begin{center}
\begin{tabular}{|l|c|c|c|}
\hline & AUC-ROC & AUC-PR & F1 score \\
\hline
2D \ac{cnn} \cite{boonyakitanont_comparison_2019} & $87.27\pm0.05$ & $72.05\pm11.46$ & $85.58\pm1.63$ \\
\hline
2D \ac{cnn} LOO \cite{gomez_automatic_2020} & N/A & N/A & $46.6 \pm 31.0$ \\
\hline
Our 1D CNN & $89.53\pm0.04$ & $74.53\pm10.56$ & $89.22\pm2.36$ \\
\hline
Our 1D CNN-FG &  $90.23\pm0.05$ & {\bf $76.73\pm11.44$} &  {$90.55\pm3.84$} \\
\hline
Our 1D CNN-GRU & {\bf $90.56\pm0.03$} & {\bf $76.77\pm8.05$} & $90.42\pm3.15$ \\
\hline
\end{tabular}
\end{center}
\end{table}

\begin{figure}
\vspace{-0.5cm}
    \centering
    \includegraphics[width=2.9in]{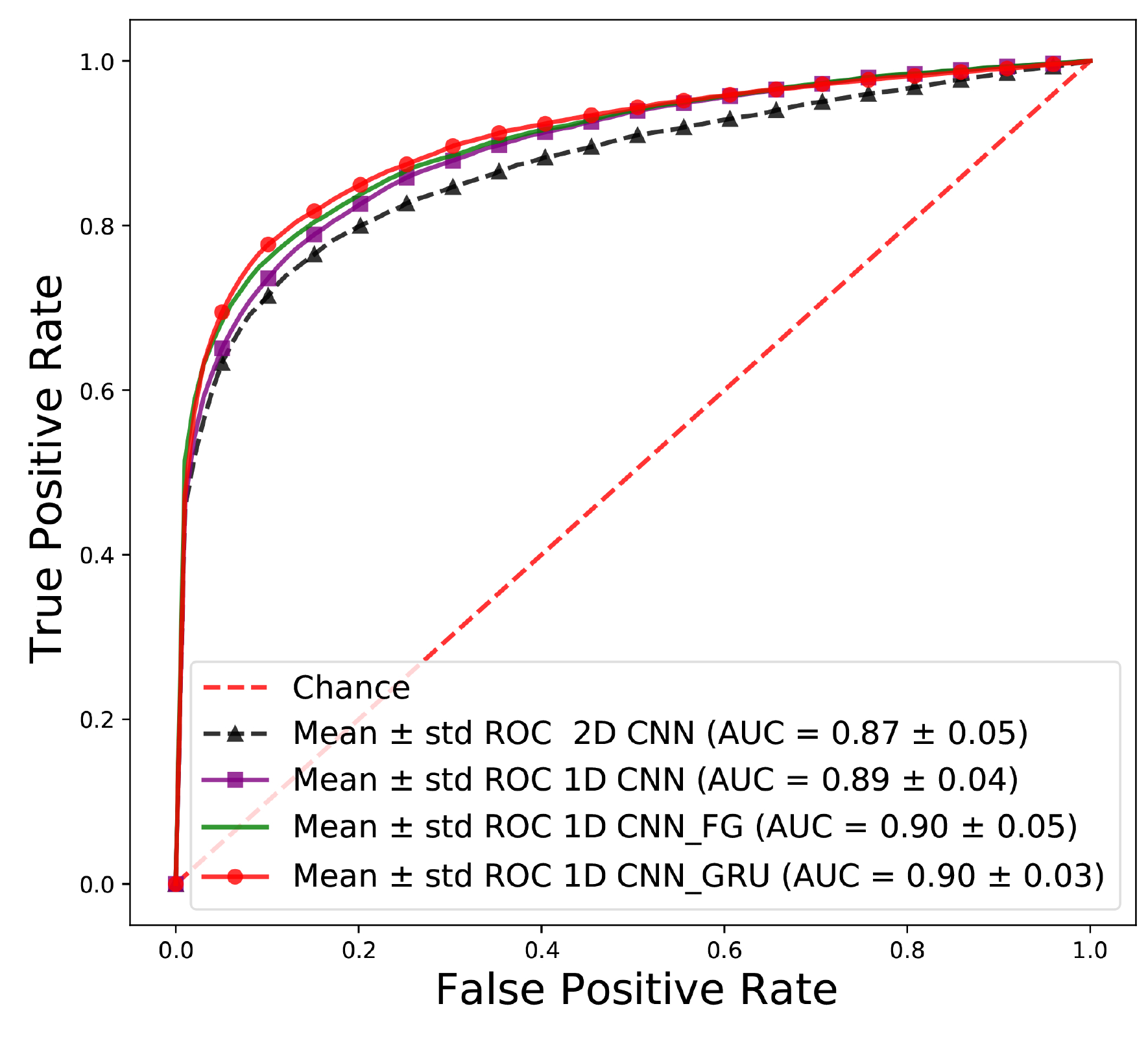}
    \vspace{-0.2cm}
    \caption{Area under ROC curve for all architectures.}
    \vspace{-0.5cm}
    \label{fig:AUC_ROC_Final}
\end{figure}


\subsection{Numerical Results}
We consider 5 models for comparison. The baselines models from \cite{boonyakitanont_comparison_2019, gomez_automatic_2020}, the 1D \ac{cnn} we proposed, the CNN-aided factor graph, and the 1D \ac{cnn} with GRU for capturing the temporal correlations. From the baseline models, we implement the exact architecture in \cite{boonyakitanont_comparison_2019} and evaluate it using the 6 fold leave-4-patient-out. For \cite{gomez_automatic_2020}, since it uses images as features, we just report the results of their leave-one-patient-out (LOO) evaluations from their paper.

Fig.~\ref{fig:AUC_ROC_Final} shows the average AUC-ROC across the 6 folds, while Table~\ref{tab:summary} summarizes all the results. As can be seen, changing the model from 2D CNN proposed in \cite{boonyakitanont_comparison_2019} to our proposed 1D \ac{cnn} architecture results in approximately 2\% improvement across all performance measure. Our hybrid model-based/data-driven \ac{cnn}-aided factor graph further improves the results by as much as 2\%. Compared to \cite{gomez_automatic_2020}, we perform a leave-4-patient-out evaluation as opposed to leave-one-patient-out. This reduces the number of patients we use for training compared to \cite{gomez_automatic_2020}. Despite this reduction, we achieve a much higher F1 score compared to \cite{gomez_automatic_2020} as shown in Table~\ref{tab:summary}. Moreover, while AUC-PR is not reported in \cite{gomez_automatic_2020}, the precision of $51.4 \pm 34.1$ and recall of $53.1 \pm 25.5$ is reported, which is significantly lower than the performance achieved by our proposed approach.

The 1D \ac{cnn} with GRU achieves the same performance as our proposed \ac{cnn}-aided factor graph inference method for all performance metrics. This demonstrates that the hybrid model-based/data-driven approach proposed here can achieve similar performance as a purely data-driven methods that employ deep and highly parameterized neural networks. This performance is achieved by our \ac{cnn}-aided factor graph at a fraction of the computational complexity of deep learning based approaches as summarized in Table~\ref{tab:numbers}.  This makes the proposed method suitable for real-time seizure~detection.

\vspace{-0.1cm}
\section{Conclusions}
\label{sec:conclusion}
\vspace{-0.1cm}
In this paper, we proposed a computationally efficient hybrid model-based/data-driven method using \ac{cnn}-aided factor graphs for seizure detection. First, we carefully designed a 1D CNN for estimating the  probability that a 4-second block of EEG recording  is a seizure block. We then used this neural network in the factor node of the factor graph for inference. We demonstrated that the proposed method generalizes well to other patients using a 6-fold leave-4-patient-out cross validation. We also showed that our algorithm achieves up to 5\% improvement in performance compared to prior work, while maintaining much lower computational complexity. This makes our approach ideal for real-time seizure detection. For future work, we plan to expand our approach to classifying focal and generalized seizures since this is a challenging task in clinical procedures. 


\vspace{-0.1cm}
\bibliographystyle{ieeetr}
\bibliography{IEEEabrv,Ref}

\begin{thebibliography}{10}

\bibitem{noauthor_epilepsy_nodate}
``Epilepsy.'' \url{https://www.who.int/news-room/fact-sheets/detail/epilepsy}.

\bibitem{park_epileptic_2018}
C.~Park, G.~Choi, J.~Kim, S.~Kim, T.-J. Kim, K.~Min, K.-Y. Jung, and J.~Chong,
  ``Epileptic seizure detection for multi-channel {EEG} with deep convolutional
  neural network,'' in {\em International Conference on Electronics,
  Information, and Communication (ICEIC)}, 2018.

\bibitem{zazzaro_eeg_2019}
G.~Zazzaro, S.~Cuomo, A.~Martone, R.~V. Montaquila, G.~Toraldo, and L.~Pavone,
  ``{EEG} signal analysis for epileptic seizures detection by applying data
  mining techniques,'' {\em Internet of Things}, vol.~14, p.~100048, 2021.

\bibitem{kulaseharan_identifying_2019}
S.~Kulaseharan, A.~Aminpour, M.~Ebrahimi, and E.~Widjaja, ``Identifying lesions
  in paediatric epilepsy using morphometric and textural analysis of magnetic
  resonance images,'' {\em NeuroImage: Clinical}, vol.~21, 2019.

\bibitem{van_klink_simultaneous_2019}
N.~van Klink, A.~Mooij, G.~Huiskamp, C.~Ferrier, K.~Braun, A.~Hillebrand, and
  M.~Zijlmans, ``Simultaneous {MEG} and {EEG} to detect ripples in people with
  focal epilepsy,'' {\em Clinical Neurophysiology}, vol.~130, no.~7,
  pp.~1175--1183, 2019.

\bibitem{pianou_imaging_2019}
N.~Pianou and S.~Chatziioannou, ``Imaging with {PET}/{CT} in {Patients} with
  {Epilepsy},'' in {\em Epilepsy {Surgery} and {Intrinsic} {Brain} {Tumor}
  {Surgery}}, pp.~45--50, Springer, 2019.

\bibitem{subasi_epileptic_2019}
A.~Subasi, J.~Kevric, and M.~Abdullah~Canbaz, ``Epileptic seizure detection
  using hybrid machine learning methods,'' {\em Neural Computing and
  Applications}, vol.~31, pp.~317--325, Jan. 2019.

\bibitem{golmohammadi_deep_2017}
M.~Golmohammadi, S.~Ziyabari, V.~Shah, S.~L. de~Diego, I.~Obeid, and J.~Picone,
  ``Deep architectures for automated seizure detection in scalp {EEGs},'' {\em
  arXiv preprint arXiv:1712.09776}, 2017.

\bibitem{slimen_eeg_2020}
I.~B. Slimen, L.~Boubchir, Z.~Mbarki, and H.~Seddik, ``{EEG} epileptic seizure
  detection and classification based on dual-tree complex wavelet transform and
  machine learning algorithms,'' {\em Journal of biomedical research}, vol.~34,
  no.~3, p.~151, 2020.

\bibitem{ahmad_prediction_2020}
S.~R.~R. Ahmad, S.~M. Sayeed, Z.~Ahmed, N.~M. Siddique, and M.~Z. Parvez,
  ``Prediction of epileptic seizures using support vector machine and
  regularization,'' in {\em 2020 IEEE Region 10 Symposium (TENSYMP)},
  pp.~1217--1220, IEEE, 2020.

\bibitem{raghu_automated_2020}
S.~Raghu, N.~Sriraam, S.~V. Rao, A.~S. Hegde, and P.~L. Kubben, ``Automated
  detection of epileptic seizures using successive decomposition index and
  support vector machine classifier in long-term {EEG},'' {\em Neural Computing
  and Applications}, vol.~32, no.~13, pp.~8965--8984, 2020.

\bibitem{li_seizure_2021}
C.~Li, W.~Zhou, G.~Liu, Y.~Zhang, M.~Geng, Z.~Liu, S.~Wang, and W.~Shang,
  ``Seizure {Onset} {Detection} {Using} {Empirical} {Mode} {Decomposition} and
  {Common} {Spatial} {Pattern},'' {\em IEEE Transactions on Neural Systems and
  Rehabilitation Engineering}, vol.~29, pp.~458--467, 2021.

\bibitem{khalilpour_application_2020}
S.~Khalilpour, A.~Ranjbar, M.~B. Menhaj, and A.~Sandooghdar, ``Application of
  {1-D CNN} to predict epileptic seizures using {EEG} records,'' in {\em 2020
  6th International Conference on Web Research (ICWR)}, pp.~314--318, IEEE,
  2020.

\bibitem{jana_1d-cnn-spectrogram_2020}
G.~C. Jana, R.~Sharma, and A.~Agrawal, ``A {1D}-{CNN}-spectrogram based
  approach for seizure detection from {EEG} signal,'' {\em Procedia Computer
  Science}, vol.~167, pp.~403--412, 2020.
\newblock Publisher: Elsevier.

\bibitem{sharan_epileptic_2020}
R.~V. Sharan and S.~Berkovsky, ``Epileptic seizure detection using
  multi-channel {EEG} wavelet power spectra and 1-{D} convolutional neural
  networks,'' in {\em 2020 42nd Annual International Conference of the IEEE
  Engineering in Medicine \& Biology Society (EMBC)}, pp.~545--548, IEEE, 2020.

\bibitem{boonyakitanont_comparison_2019}
P.~Boonyakitanont, A.~Lek-uthai, K.~Chomtho, and J.~Songsiri, ``A {Comparison}
  of {Deep} {Neural} {Networks} for {Seizure} {Detection} in {EEG} {Signals},''
  {\em bioRxiv}, p.~702654, 2019.

\bibitem{roy18GRU}
S.~{Roy}, I.~{Kiral-Kornek}, and S.~{Harrer}, ``Deep learning enabled automatic
  abnormal {EEG} identification,'' in {\em International Conference of the IEEE
  Engineering in Medicine and Biology Society (EMBC)}, pp.~2756--2759, 2018.

\bibitem{liang_scalp_2020}
W.~Liang, H.~Pei, Q.~Cai, and Y.~Wang, ``Scalp {EEG} epileptogenic zone
  recognition and localization based on long-term recurrent convolutional
  network,'' {\em Neurocomputing}, vol.~396, pp.~569--576, 2020.

\bibitem{lee_classification_2018}
M.~Lee, I.~Youn, J.~Ryu, and D.-H. Kim, ``Classification of both seizure and
  non-seizure based on {EEG} signals using hidden {Markov} model,'' in {\em
  2018 IEEE International Conference on Big Data and Smart Computing
  (BigComp)}, pp.~469--474, IEEE, 2018.

\bibitem{loeliger2004introduction}
H.-A. Loeliger, ``An introduction to factor graphs,'' {\em {IEEE} Signal
  Process. Mag.}, vol.~21, no.~1, pp.~28--41, 2004.

\bibitem{emmady_eeg_2020}
P.~D. Emmady and A.~C. Anilkumar, ``{EEG}, {Abnormal} {Waveforms},'' {\em
  StatPearls [Internet]}, 2020.

\bibitem{goldberger_physiobank_2000}
A.~L. Goldberger, L.~A. Amaral, L.~Glass, J.~M. Hausdorff, P.~C. Ivanov, R.~G.
  Mark, J.~E. Mietus, G.~B. Moody, C.-K. Peng, and H.~E. Stanley,
  ``{PhysioBank}, {PhysioToolkit}, and {PhysioNet}: components of a new
  research resource for complex physiologic signals,'' {\em circulation},
  vol.~101, no.~23, pp.~e215--e220, 2000.

\bibitem{klonowski2009everything}
W.~Klonowski, ``Everything you wanted to ask about eeg but were afraid to get
  the right answer,'' {\em Nonlinear biomedical physics}, vol.~3, no.~1,
  pp.~1--5, 2009.

\bibitem{gomez_automatic_2020}
C.~Gomez, P.~Arbelaez, M.~Navarrete, C.~Alvarado-Rojas, M.~Le~Van~Quyen, and
  M.~Valderrama, ``Automatic seizure detection based on imaged-{EEG} signals
  through fully convolutional networks,'' {\em Scientific Reports}, vol.~10,
  p.~21833, Dec. 2020.

\bibitem{cho_comparison_2020}
K.-O. Cho and H.-J. Jang, ``Comparison of different input modalities and
  network structures for deep learning-based seizure detection,'' {\em
  Scientific Reports}, vol.~10, p.~122, Jan. 2020.

\bibitem{kschischang2001factor}
F.~R. Kschischang, B.~J. Frey, and H.-A. Loeliger, ``Factor graphs and the
  sum-product algorithm,'' {\em {IEEE} Trans. Inf. Theory}, vol.~47, no.~2,
  pp.~498--519, 2001.

\bibitem{forney}
G.~D. Forney, ``Codes on graphs: Normal realizations,'' {\em IEEE Trans. Inf.
  Theory}, vol.~47, pp.~520--548, 2001.

\bibitem{mormann_epileptic_2003}
F.~Mormann, T.~Kreuz, R.~G. Andrzejak, P.~David, K.~Lehnertz, and C.~E. Elger,
  ``Epileptic seizures are preceded by a decrease in synchronization,'' {\em
  Epilepsy research}, vol.~53, no.~3, pp.~173--185, 2003.

\bibitem{shlezinger2020inference}
N.~Shlezinger, N.~Farsad, Y.~C. Eldar, and A.~J. Goldsmith, ``Learned factor
  graphs for inference from stationary time sequences,'' {\em arXiv preprint
  arXiv:2006.03258}, 2020.

\bibitem{shlezinger2020viterbinet}
N.~Shlezinger, N.~Farsad, Y.~C. Eldar, and A.~J. Goldsmith, ``{ViterbiNet}: A
  deep learning based {Viterbi} algorithm for symbol detection,'' {\em IEEE
  Trans. Wireless Commun.}, vol.~19, no.~5, pp.~3319--3331, 2020.

\bibitem{powers_evaluation_2020}
D.~M. Powers, ``Evaluation: from precision, recall and {F}-measure to {ROC},
  informedness, markedness and correlation,'' {\em arXiv preprint
  arXiv:2010.16061}, 2020.

\end{thebibliography}

\end{document}